\documentclass[12pt]{article}

\usepackage{graphics,amssymb,float}
\usepackage{graphicx}
\usepackage{caption}
\usepackage{rotating}
\usepackage{dcolumn}
\usepackage{bm}
\usepackage{cite}
\usepackage{amsmath}
\usepackage{indentfirst} 
\usepackage{epstopdf}
\usepackage[usenames,dvipsnames]{xcolor}
\usepackage{setspace}
\usepackage{hhline}
\usepackage{enumitem}

\usepackage{xcolor}
\usepackage{color}
\usepackage{colortbl}

\makeatletter

\@addtoreset{equation}{section}
\makeatother

\def\lsim{\;\raise0.3ex\hbox{$<$\kern-0.75em\raise-1.1ex\hbox{$\sim$}}\;}
\def\gsim{\;\raise0.3ex\hbox{$>$\kern-0.75em\raise-1.1ex\hbox{$\sim$}}\;}
\def\beq{\begin{equation}}   \def\eeq{\end{equation}}
\def\ba{\begin{array}}       \def\ea{\end{array}}
\def\bea{\begin{eqnarray}}   \def\eea{\end{eqnarray}}

\def\nl{\newline}

\def\l{\lambda}

\def\t{\theta}

\begin{document}

\begin{titlepage}


\begin{center}
\vspace{1cm}

{\Large\bf NMSSM Explanation for Excesses in the Search for \\[0.25cm] 
Neutralinos and Charginos and a 95 GeV Higgs Boson}

\vspace{2cm}

{\bf{Ulrich Ellwanger$^a$, Cyril Hugonie$^b$, Stephen F. King$^c$, Stefano Moretti$^{c,d}$}}\\
\vspace{1cm}
\it $^a$ IJCLab,  CNRS/IN2P3, University  Paris-Saclay, 91405  Orsay,  France\\
\it $^b$ LUPM, UMR 5299, CNRS/IN2P3, Universit\'e de Montpellier, 34095 Montpellier, France\\
\it $^c$ School of Physics and Astronomy, University of Southampton, Highfield,
Southampton, SO17 1BJ, United Kingdom
\\
\it $^d$ Department of Physics \& Astronomy, Uppsala University, Box 516, 75120 Uppsala, Sweden

\end{center}
\vspace{2cm}

\begin{abstract}

The observed excesses in the search for neutralinos and charginos by ATLAS and CMS can be fitted simultaneously in the minimal supersymmetric standard model (MSSM) assuming a light higgsino mass,
of magnitude less than about 250 GeV, and a {\em compressed} higgsino dominated neutralino and chargino spectrum, with
$5-10\%$ mass splittings.
However, light higgsinos as dark matter would have far too large direct detection cross sections. We consider the next-to-MSSM (NMSSM) with an additional singlino-like lightest supersymmetric particle (LSP) a few GeV below the next-to-lightest supersymmetric particle (NLSP). Sparticles prefer to decay first into the NLSP and remnants from the final decay into the LSP are too soft to contribute to the observed signals. Co-annihilation in the higgsino-sector can generate a relic density in the WMAP/Planck window. The singlino-like LSP has automatically a direct detection cross section below present and future sensitivities: a direct detection signal in the near future would exclude this scenario. The singlet-like Higgs scalar of the NMSSM can have a mass around 95~GeV and signal cross sections in the $b\bar{b}$ channel at LEP and in the $\gamma\gamma$ channel at the LHC compatible with the respective observations.

\end{abstract}

\end{titlepage}

\section{Introduction}

One of the most important tasks of the ATLAS and CMS experiments at the LHC is the search for deviations from predictions of the Standard Model (SM), which could be signs for physics beyond the Standard Model (BSM). Particularly relevant are such deviations when observed simultaneously by both experiments. 

Signals for supersymmetry (SUSY) extensions of the SM include missing transverse energy ($E_T^\text{miss}$), if R-parity is conserved and the Lightest Supersymmetric Particle (LSP) is neutral and stable. Often the LSP is the lightest  neutralino\footnote{Neutralinos (and charginos) are the spin-1/2 superparticle mass eigenstates which are linear combinations of superpartners of electroweak gauge bosons (gauginos) and Higgs bosons (Higgsinos), they are present in all supersymmetric extensions of the SM.}. In the minimal supersymmetric standard model (MSSM), the lightest neutralino is a candidate for the LSP and, being stable, is a welcome candidate for dark matter (DM). The decays of heavier neutralinos and charginos into the LSP can proceed via the emission of $Z$ or $W$ bosons, which can be detected via their leptonic decays. Hence, interesting search channels are final states with $E_T^\text{miss}$ and leptons which can be signals for the pair production of neutralinos and charginos.

After the end of Run~2 of the LHC, both ATLAS and CMS have published their corresponding SUSY search results in leptonic channels in \cite{ATLAS:2019lng,ATLAS:2021moa,ATLAS:2023act,ATLAS:2024qxh} and \cite{CMS:2020bfa,CMS:2021cox,CMS:2021edw,CMS:2024gyw}, respectively.
Interestingly, both experiments have observed (mild) excesses of events with respect to the SM. These extra events correspond to final states expected from chargino-neutralino pair production in the case where chargino and neutralino decays proceed via off-shell $W$ and $Z$ bosons leading to soft leptons. 

The precise kinematic properties of these final states depend on the spectra of the participating neutralinos and charginos. If the corresponding charginos are dominantly superpartners of $W^\pm$-bosons, one expects a neutralino of similar mass, the superpartner of the $W^3$-boson before electroweak mixing. A candidate for the LSP is then a bino, the superpartner of the $B$-boson before electroweak mixing.
In contrast, if the chargino is dominantly the superpartner of a charged Higgs boson (present in all supersymmetric extensions of the SM), one expects two light neutralinos (dominantly higgsinos), one heavier and one lighter than the lightest chargino (dominantly higgsino), whose mass lies approximately halfway between the two light neutralino masses, since all three lightest ``ino'' states are dominantly arising from the higgsino triplet (two neutral Majorana higgsinos and one charged Dirac higgsino). These two scenarios are denoted by wino/bino or higgsino scenario, respectively, by the experimental collaborations. (The wino/bino scenarios depend somewhat on the relative sign among the neutral wino and bino masses, and are denoted by wino/bino+ and wino/bino- correspondingly.)

The ATLAS and CMS collaborations interpret the above observed events in both wino/bino and higgsino scenarios and derive observed limits on the neutralino/chargino masses in both scenarios. Interestingly, due to a mild excess of events both collaborations obtain observed limits below the expected ones, although only at the $(1-2)\,\sigma$ level. 
An interpretation of the deviations from the SM depends on the assumed masses in the neutralino/chargino sector and on the assumed scenario. 
Within the wino/bino scenario the preferred value (corresponding to an excess beyond $1\,\sigma$) for the chargino-neutralino mass difference $\Delta m$ is $\Delta m < 25$~GeV in case of the ATLAS results, but $\Delta m > 25$~GeV in case of the CMS results. If the higgsino scenario is assumed, a value for the neutral heavy-light mass difference $\Delta m$ in the $15-30$~GeV range explains simultaneously both ATLAS and CMS excesses of events beyond the $1\,\sigma$ level.

The excesses of events have been interpreted in the framework of the MSSM in the context of both wino/bino and higgsino scenarios in \cite{Chakraborti:2024pdn}. In their Fig.~1 the authors of \cite{Chakraborti:2024pdn} compare the resulting limits from ATLAS in \cite{ATLAS:2021moa} shown in Fig.~16d and from CMS in \cite{CMS:2021edw} shown in Fig.~9a, which leads to the above conclusions. The authors of \cite{Chakraborti:2024pdn} apply an experimental mass resolution of $\sim 5$~GeV; then a common region of chargino-neutralino mass differences describing simultaneously minor excesses in both ATLAS and CMS data exists also for the wino/bino scenario. 

In the MSSM, the higgsino scenario is ruled out since DM in the form of light higgsino LSPs as required here would have too large direct detection cross sections in order to comply with the latest constraints from the LZ experiment \cite{LZ:2022lsv}. However, these constraints are easy to satisfy within the Next-to-MSSM  (NMSSM)~\cite{Ellwanger:2009dp}.
The NMSSM contains two extra singlet-like Higgs bosons (a scalar and a pseudo-scalar) as well as an extra singlet-like neutralino - the singlino - which can well be the LSP.
Singlino DM can have very small direct detection cross sections, but can simultaneously generate a relic density compatible with the WMAP/Planck value $\Omega_{\rm DM} h^2 = 0.1187$ \cite{Hinshaw:2012aka,Ade:2013zuv}. For these reasons the singlino has become an attractive DM candidate, at least after the latest constraints from the LZ experiment 
\cite{Domingo:2022pde,Cao:2022ovk,Wang:2023suf,Cao:2023juc,Heng:2023wqf,Ellwanger:2023zjc,Cao:2023gkc,Cao:2024axg,Ellwanger:2024txc}.
In the context of excess events in searches for neutralinos and charginos, the NMSSM with a singlino-like LSP has already been considered in \cite{Agin:2024yfs}. There, the authors considered simultaneously excesses of events with monojets \cite{Agin:2023yoq}.

The experimental signatures for supersymmetric extensions of the SM can remain nearly unchanged relative to the MSSM in the presence of the singlino of the NMSSM with its possibly very small couplings to the fermions and bosons of the MSSM. Then the MSSM-like superpartners of the SM do not decay directly into the singlino, unless this is the only decay channel allowed by the conservation of R-parity. Hence the decay chains of superpartners of the SM have the form of the MSSM, and terminate provisionally with the Next-to-LSP (NLSP). Only subsequently, the NLSP decays in a last step into the singlino LSP plus a photon or an off-shell $Z$ boson. The NLSP can well be the lighter neutral higgsino whose presence is assumed in the higgsino scenario which can explain simultaneously both ATLAS and CMS excesses of events. If the singlino is just a few GeV lighter than the higgsino NLSP, the decay products of the higgsino-to-singlino decay will remain too soft to be observable at the LHC. 

In this paper, then, we shall interpret the ATLAS and CMS SUSY search excesses in the framework of the NMSSM, assuming the above scenario, namely the ``compressed'' light higgsino-like triplet consisting of  ($\chi_3^0$, $\chi_1^{\pm}$, $\chi_2^0$)  plus a nearby singlino LSP $\chi_1^0$, with all masses less than about 250 GeV with of order $5-10\%$ mass splittings. 
Note that this NMSSM scenario differs from the one considered in \cite{Agin:2024yfs}: there the higgsino-singlino mass splitting is assumed to be in the $5-20$ GeV range, and the decay products of the higgsino-to-singlino decay are part of the signal. Consequently the statements in \cite{Agin:2024yfs} on the NMSSM in connection with events with monojets are not applicable.

Another attractive feature of the NMSSM is the extra dominantly gauge singlet scalar Higgs state which can naturally have a mass in the 95~GeV range. Hints for such an additional Higgs boson have been observed by the CMS and ATLAS experiments. A search at Run~1 by CMS showed a $\sim 2\, \sigma$ excess at 97~GeV \cite{CMS-PAS-HIG-14-037}, which was confirmed by CMS later in \cite{CMS:2018cyk} and in \cite{CMS-PAS-HIG-20-002} for a mass hypothesis of 95.4~GeV. A somewhat less sensitive search in the diphoton channel by ATLAS in \cite{ATLAS-CONF-2018-025} led to an upper limit on the fiducial cross section which did not contradict the possible excess observed by CMS, but a recent analysis by ATLAS in the diphoton channel in \cite{ATLAS-CONF-2023-035} showed a mild excess of $1.7\, \sigma$ at 95~GeV.
Actually already the combination of searches for the SM Higgs boson by the ALEPH, DELPHI, L3 and OPAL experiments at LEP \cite{LEPWorkingGroupforHiggsbosonsearches:2003ing} showed some mild excess of events in the $Z^*\to Z+b\bar{b}$ channel in the mass region 95~GeV. The fact that such a state can well be described within the NMSSM has already been widely discussed \cite{Belanger:2012tt,Cao:2016uwt,Biekotter:2017xmf,Hollik:2018yek,Domingo:2018uim,Wang:2018vxp,Cao:2019ofo,Choi:2019yrv,Hollik:2020plc,Biekotter:2021qbc,Adhikary:2022pni,Li:2022etb,Biekotter:2023oen,Ellwanger:2023zjc,Cao:2023gkc,Li:2023kbf,Roy:2024yoh}.

In   Section~2 we present our conventions for the neutralino/chargino sector of the NMSSM. In Section~3 we
discuss the constraints that we will apply to the parameter space of the NMSSM, which we scan employing the codes \texttt{NMSSMTools-6.0.3} \cite{Ellwanger:2004xm,Ellwanger:2005dv,NMSSMTools} and {\sf micrOMEGAs$\_$3}~\cite{Belanger:2013oya}. The results will be presented in the form of figures and benchmark points in Section~4. Then, we conclude with a summary in Section~5.

\section{The Neutralino/Chargino Sector of the NMSSM}

We consider the ${\mathbb Z}_3$ invariant NMSSM with the superpotential
\beq\label{eq:2.1}
W_\text{NMSSM} = \lambda \hat S \hat H_u\cdot \hat H_d + \frac{\kappa}{3} 
\hat S^3 +\dots
\eeq
where the dots denote the Yukawa couplings of the superfields $\hat H_u$ and $\hat H_d$
to the quarks and leptons as in the MSSM. Once the scalar component of the superfield
$\hat S$ develops a vacuum expectation value  $\left< S\right>\equiv s$, the first term in
$W_\text{NMSSM}$ generates an effective $\mu$-term with
\beq\label{eq:2.2}
\mu_\mathrm{eff}=\lambda\, s\; .
\eeq
(Subsequently the index $_\mathrm{eff}$ of $\mu$ will be omitted for
simplicity.) Hence, $\mu$ generates Dirac mass terms for the charged and
neutral SU(2) doublet higgsinos $\psi_u$ and $\psi_d$.

In the ``decoupling'' limit $\lambda, \kappa \to 0$ all components of the superfield
$\hat S$ decouple from all components of $\hat H_u,\ \hat H_d$ and the matter
superfields. However, since $s\sim M_{\rm SUSY}/\kappa$ where $M_{\rm SUSY}$ denotes
the scale of soft Susy breaking masses and trilinear couplings, 
$\mu$ remains of ${\cal O}(M_{\rm SUSY})$ in the decoupling limit
provided $\lambda/\kappa\sim {\cal O}(1)$.

Including bino ($\widetilde{B}$) masses $M_1$ and wino ($\widetilde{W}^3$) masses
$M_2$, the symmetric $5 \times 5$ neutralino mass matrix ${\cal M}_0$
in the basis $\psi^0 = (-i\widetilde{B} ,
-i\widetilde{W}^3, \psi_d^0, \psi_u^0, \psi_S)$
is given by \cite{Ellwanger:2009dp}
\beq\label{eq:2.3}
{\cal M}_0 =
\left( \ba{ccccc}
M_1 & 0 & -\frac{g_1 v_d}{\sqrt{2}} & \frac{g_1 v_u}{\sqrt{2}} & 0 \\
& M_2 & \frac{g_2 v_d}{\sqrt{2}} & -\frac{g_2 v_u}{\sqrt{2}} & 0 \\
& & 0 & -\mu & -\lambda v_u \\
& & & 0 & -\lambda v_d \\
& & & & 2\kappa s
\ea \right),
\eeq
where $v_u^2 + v_d^2 =v^2 \simeq (174\ \text{GeV})^2$ and $\frac{v_u}{v_d}=\tan\beta$.
The eigenvalues of ${\cal M}_0$ will be denoted by $M_{\chi_i^0}$, the eigenstates by $\chi_i^0$, $i=1...5$ ordered in increasing mass. The lightest eigenstate $\chi_1^0$ is assumed to represent the LSP (which could, in principle, also be represented by a lighter sneutrino).

The charged $SU(2)$ gauginos are
$\l^- = \frac{1}{\sqrt{2}}\left(\l_2^1 + i \l_2^2\right)$, and
$\l^+ = \frac{1}{\sqrt{2}}\left(\l_2^1 - i \l_2^2\right)$, which mix
with the charged higgsinos $\psi_u^+$ and $\psi_d^-$. Defining 
\beq\label{2.4}
\psi^+ = \left(\ba{c} -i\l^+ \\ \psi_u^+ \ea\right)\,, \qquad
\psi^- =  \left(\ba{c} -i\l^- \\ \psi_d^- \ea\right)\,,
\eeq
the corresponding mass terms in the Lagrangian can be written as
\beq\label{2.5}
{\cal L} = -\frac{1}{2} (\psi^+ , \psi^-) \left(\ba{cc} 0 & X^T \\ 
X & 0 \ea\right) \left(\ba{c} \psi^+ \\ \psi^- \ea\right) + \mathrm{h.c.} 
\eeq
with
\beq\label{2.6}
X = \left(\ba{cc} M_2 & g_2 v_u \\ g_2 v_d & \mu \ea\right)\,.
\eeq

The diagonalization of the (not symmetric) $2 \times 2$ chargino mass
matrix $X$ (\ref{2.6}) in the basis $\psi^-$, $\psi^+$ requires different rotations of $\psi^-$
and $\psi^+$ into the 2-component mass eigenstates $\chi^-$, $\chi^+$ as
\beq\label{2.7}
\chi^- = U \psi^- , \qquad \chi^+ = V \psi^+
\eeq
with
\beq\label{2.8}
U = \left(\ba{cc} \cos\t_U & \sin\t_U \\ -\sin\t_U & \cos\t_U\ea \right)\,, \qquad
V = \left(\ba{cc} \cos\t_V & \sin\t_V \\ -\sin\t_V & \cos\t_V \ea \right)\,.
\eeq

The eigenvalues of the chargino mass matrix $X$ are $M_{\chi_i^\pm}$ and the eigenstates are $\chi_i^\pm$, $i=1...2$ ordered in increasing mass. 

\section{Constraints on the Parameter Space of the NMSSM}

We scan the parameter space over the NMSSM-parameters $\lambda$, $\kappa$, $A_\lambda$, $A_\kappa$, $\mu_{\text{eff}}$, $\tan\beta$ 
in the Higgs sector and non universal soft SUSY breaking terms in the gaugino/sfermion sector; for the definitions of the latter we refer to the review in \cite{Ellwanger:2009dp}. All soft SUSY breaking terms are taken below 5~TeV. The aim is to reproduce 
the ``compressed'' light higgsino-like spectrum consisting of ($\chi_3^0$, $\chi_1^{\pm}$, $\chi_2^0$)  plus the nearby singlino LSP $\chi_1^0$.

Of relevance for a potential description of the excesses observed by ATLAS and CMS are the masses of the charged and neutral higgsino-like states as well as of the singlino LSP. Since we focus on the higgsino scenario, we assume $M_1, M_2 > \mu$ leading to heavier binos and winos than higgsinos. The mass of the singlino is approximately given by $2\kappa s$ and we assume $\lambda \sim {\cal O}(10^{-2})$ such that its mixings and couplings to other particles are small. Its mass is assumed just (less than 5~GeV) below the mass of the lighter neutral higgsino such that it corresponds to the LSP $\chi_1^0$ while  the neutral higgsinos are $\chi_2^0$ and $\chi_3^0$. 

The mass splittings $M_{\chi_3^0} - M_{\chi_1^\pm}$ and $M_{\chi_1^\pm} - M_{\chi_2^0}$ among the charged and neutral higgsinos are not exactly the same as assumed by the experimental collaborations for the interpretation of excesses within the higgsino-scenario. In order to reproduce these assumptions within the experimental mass resolution we require 
\beq\label{delta_mneu}
|M_{\chi_1^\pm}-(M_{\chi_3^0} + M_{\chi_2^0})/2| < 5\,\text{GeV}\,.
\eeq
As discussed in the Introduction, a simultaneous fit to both excesses of events observed by ATLAS and CMS is possible, more precisely assuming 
\beq\label{delta_m}
15\, \text{GeV} < \Delta m < 30\,\text{GeV}\,, \qquad \Delta m = M_{\chi_3^0} - M_{\chi_2^0},
\eeq
which we impose on our parameters, resulting in a  ``compressed'' light higgsino-like triplet of states ($\chi_3^0$, $\chi_1^{\pm}$, $\chi_2^0$).
In addition we will have a nearby singlino LSP $\chi_1^0$ with a mass a few GeV below the one of $\chi_2^0$.
In order to explain the excesses we require all these masses to be less than about 250 GeV, as we now discuss.

The signal cross sections required to reproduce the excesses are estimated as follows. First we employ the code \texttt{Prospino2} at NLO \cite{Beenakker:1999xh} in order to compute the higgsino production cross sections for the various relevant neutral and charged higgsino-like masses $M_{\chi_3^0}$ and $M_{\chi_1^\pm}$. Assuming 100\% branching fractions for the decays into $\chi_2^0$ plus $Z^*$ and $W^*$ bosons for the masses corresponding to the expected limits, we obtain the expected upper limits on the higgsino pair production cross sections. 
Comparing to the observed limits on the relevant charged higgsino masses, we obtain the observed upper limits on the higgsino-like pair production cross sections. The difference between the expected and observed upper limits on the production cross sections gives an estimate for the required signal cross sections. For the considered range of $\Delta m$ and within the uncertainties for the expected cross sections we obtain estimates for the required signal cross sections of $0.2-1$~pb, which coincide the estimates in \cite{Chakraborti:2024pdn}.

\texttt{Prospino2} at NLO is also used to compute the higgsino-like pair production cross sections for the relevant neutralino and chargino masses $M_{\chi_3^0}$ and $M_{\chi_1^\pm}$ within the NMSSM. These are multiplied by the branching fractions for the decays into $\chi_2^0$ plus $Z^*$ and $W^*$ bosons using the code {\sf NMSDECAY}~\cite{Das:2011dg} which lead to the effective signal cross section $\sigma_\text{eff}^\text{signal}$\footnote{{\sf NMSDECAY} is based on the code {\sf SDECAY}~\cite{Muhlleitner:2003vg}, and for loop induced neutralino decays into photons {\sf SDECAY} uses the conventions for signs for chargino masses and the matrices $U$ and $V$ in eq.~(\ref{2.8}) from \cite{Haber:1988px}. These were not implemented in earlier versions of {\sf NMSSMTools}, hence an updated version \texttt{NMSSMTools-6.0.3} has been employed.}.

The sum over the final decays $\chi_2^0 \to \chi_1^0 + Z^*$ or $\gamma$ proceeds with 100\% probability, but they are considered as unobservable. Thus we require
\beq\label{sigeff}
0.2\,\text{pb} < \sigma_\text{eff}^\text{signal} < 1.0\,\text{pb}\, .
\eeq

The final decays $\chi_2^0 \to \chi_1^0 + X$, $X =  Z^*/\gamma$ are unobservable if the mass difference $\varepsilon = M_{\chi_2^0} - M_{\chi_1^0}$ is small enough. In an expansion in powers of $\varepsilon$ one can derive from energy and momentum conservation
\beq
E_X^2 + ({p_T}_X)^2 \left( \frac{ {p_T}_{\chi_2^0} }{ M_{\chi_2^0} }\right)^2 < \varepsilon^2 
\eeq
which leads indeed to $E_X$ and/or ${p_T}_X$ below the applied cuts for most naturally occuring values of ${p_T}_{\chi_2^0}$ and $M_{\chi_2^0}$ if
\beq\label{dm2m1}
\varepsilon < 5\ \text{GeV}\, .
\eeq

In principle, off-shell sleptons (selectrons, smuons) can also contribute to the signal cross sections consisting in $E_T^\text{miss}$ and soft leptons. In order to estimate these contributions, dedicated simulations and comparisons with the excesses observed in the well-defined signal regions would be necessary which is beyond the scope of the present work. In order to avoid such contributions to the signal cross sections we assume all slepton masses at or beyond 1~TeV.

Herewith we conclude the applied constraints related to the potential description of the excesses observed by ATLAS and CMS.
For the DM relic density we require $\Omega_{\rm DM} h^2 = 0.1187$, allowing for $\pm 10\%$ theoretical uncertainty. Despite the weak couplings of the singlino LSP the required reduction of its relic density is easy to achieve through co-annihilation: by construction, the masses of the higgsino NLSP and chargino are not far above the mass of the LSP. Hence, at a temperature in the early universe around the LSP mass, the densities of the higgsino-like NLSP and chargino are similar to the density of the singlino LSP. The higgsino-like NLSP and chargino undergo annihilation processes which are not suppressed by small couplings and reduce this way, through exchange of their densities with the density of the LSP via equilibrium processes, also the latter. (These processes are suppressed only for couplings $\lambda \lsim 10^{-5}$.) The final LSP relic density can easily reach the required value above.

In order to make sure that the higgsino-like signal cross sections at the LHC are not spoiled by direct decays into the singlino LSP, the Yukawa coupling $\lambda$ in the neutralino mass matrix (\ref{eq:2.3}) has to be small enough, i.e. of the order $\lambda \lsim 10^{-2}$ which implies small mixing angles of the mostly singlino LSP with the neutral higgsinos. Then the LSP direct detection cross sections satisfy automatically the LZ constraints from \cite{LZ:2022lsv}.


The NMSSM also contains always an extra singlet-like scalar Higgs state whose mass depends on the yet unspecified trilinear coupling $A_\kappa$. This parameter can easily have a value such that the mass of the extra singlet-like scalar is in the 95~GeV range. We will therefore note this state $H_{95}$ and allowing for some theoretical uncertainty we require $M_{H_{95}} = 95.4 \pm 3$~GeV.
The best fits for the diphoton signal of $H_{95}$ observed by CMS and ATLAS were combined in \cite{Biekotter:2023oen}. The authors in \cite{Biekotter:2023oen} obtain
\beq\label{mugamgam}
\mu_{\gamma\gamma}^{\rm LHC} = \frac{\sigma(gg \to H_{95}\to \gamma\gamma)}{\sigma(gg \to H^{\rm SM}_{95} \to \gamma\gamma)}
=  {0.24^{+0.09}_{-0.08}}\; .
\eeq
Here $H^{\rm SM}_{95}$ denotes a SM-like Higgs boson with a mass of $\sim 95$~GeV. The $\sim 2\, \sigma$ excess at LEP was quantified in \cite{Cao:2016uwt}. Let us denote the reduced coupling of $H_{95}$ to vector bosons $W^\pm, Z$ (relative to the coupling of a SM-like Higgs boson of corresponding mass) by $C_V(H_{95})$. Then the authors in \cite{Cao:2016uwt} define
\beq\label{mulep}
\mu^{\rm LEP}_{bb} \equiv C_V(H_{95})^2\times {BR}(H_{95}\to b\bar{b})/{BR}(H^{\rm SM}_{95}\to b\bar{b})= 0.117 \pm 0.057\, .
\eeq
We require $\mu^{\rm LEP}_{bb}$ in the $2\, \sigma$ range of \eqref{mulep} and $\mu_{\gamma\gamma}^{\rm LHC}$ in the $2\, \sigma$ range of \eqref{mugamgam}.

Actually hints for $H_{95}$ have also been observed in the di-tau channel by CMS in \cite{CMS:2022goy}, where the observed excess corresponds to a cross section times branching fraction
\beq\label{ditau}
\sigma(gg \to H_{95} \to \tau\tau)= 7.8^{+3.9}_{-3.1}~\text{pb}\; . 
\eeq
Relative to $H^{\rm SM}_{95}$ we obtain
\beq\label{ditau1}
\mu_{\tau\tau}^{\rm LHC} = \frac{\sigma(gg \to H_{95}\to \tau\tau)}{\sigma(gg \to H^{\rm SM}_{95} \to \tau\tau)}
= 1.38^{+0.69}_{-0.55}\; .
\eeq
Couplings of $H_{95}$ to gluons (required for its production at the LHC), photons (required for its observation at the LHC) and $b$-quarks (required for its observation at LEP) are induced by the mixing of $H_{95}$ with the SM-like Higgs boson with mass near 125~GeV.
For the SM-like Higgs boson we require that the couplings in the $\kappa$-framework satisfy combined limits of CMS \cite{CMS:2022dwd} and ATLAS \cite{ATLAS:2022vkf}. These limits require that the reduced couplings $C_V(H_{\rm SM}$) of the SM-like Higgs boson to gauge fields satisfy $C_V(H_{\rm SM}$)~$\gsim$~0.96, hence the mixing angle between $H_{95}$ and $H_{\rm SM}$ is limited such that the reduced coupling of $H_{95}$ to $\tau\tau$ is bounded to below $25\%$. Then the (reduced) cross section $\times$ BR($H_{95}\to \tau\tau$) can hardly exceed 0.1 pb  in contrast to what is desired within the $2\, \sigma$ range for the di-tau excess $\mu_{\tau\tau}^{\rm LHC}$ in \eqref{ditau1}: its description within the $2\, \sigma$ range is therefore left aside since impossible for the type II Yukawa couplings present in the (N)MSSM to provide a suitable excess (i.e., we have to assume that the measured value represents a statistical fluctuation).

The constraint from the anomalous magnetic moment of the muon $a_\mu$ \cite{Muong-2:2006rrc,Muong-2:2021ojo} as in \cite{Domingo:2022pde} cannot be satisfied if the soft SUSY-breaking smuon, bino and wino masses are simultaneously large. According to the SM contribution to $a_\mu$ on the lattice as in \cite{Borsanyi:2020mff}, the deviation of the measurements from the SM value may even not be that large (beyond the $2\,\sigma$ level). Hence we will not consider this constraint at present.

Constraints on the sparticle spectrum from the absence of signals at the LHC are taken into account using the code \texttt{SModels-2.2.0} \cite{Kraml:2013mwa,Dutta:2018ioj,Khosa:2020zar,Alguero:2021dig}. Constraints from recent searches for light neutralinos or charginos by ATLAS and/or CMS are taken into account by construction whereas corresponding constraints from LEP are build into the code NMSSMTools.
Moreover, in the presence of several light electroweakly interacting sparticles, heavy sparticles (squarks, gluino) can undergo several distinct cascade decays such that the probability for each of them is reduced. As a consequence such scenarios are difficult to rule out.

Furthermore we impose constraints from $b$-physics and constraints from searches for BSM Higgs bosons by ATLAS and CMS as implemented in \texttt{NMSSMTools-6.0.3}. The references to constraints from BSM Higgs boson searches and $b$-physics (of little relevance here) are listed on the web page {\sf https://www.lupm.in2p3.fr/users/nmssm/history.html}. Constraints from the absence of a Landau singularity for the Yukawa couplings below the grand unification theory (GUT) scale confine values of the NMSSM-specific coupling $\lambda$ to $\lambda \lsim 0.7$, which is satisfied automatically. 
Herewith we conclude the list of applied constraints on the parameter space during the scan.

\section{Results}

In Table~1 we show the contributing range of parameters for the NMSSM specific parameters and the soft SUSY-breaking masses $M_1$, $M_2$ and $M_3$ where the latter refer to the bino, wino and gluino masses, respectively. The soft SUSY-breaking mass terms for scalars vary from 1~TeV (as imposed for sleptons) and 272~GeV (for squarks) up to 5~TeV. The upper limits of $\pm 5$~TeV are imposed by hand. Only negative values for $\mu$ and $A_\lambda$ have been found. Given the possibly complicated squark decay cascades, squarks can be relatively light without contradicting constraints from searches at the LHC. In any case, due to radiative corrections the pole squark masses can differ considerably from the soft SUSY-breaking mass terms. 
The MSSM-like CP-even, CP-odd and charged Higgs masses are in the $1.5-2.5$~TeV range.

\begin{table}[h!]
\begin{center}
\begin{tabular}{| c | c | c | c | c | c |  }
\hline
 $\lambda$  & $\kappa$  & $A_\lambda$ &$A_\kappa$ & $\mu_{\rm eff}$ & $\tan\beta$ 
\\
\hline
$0.013 - 0.019$ & $0.0058 - 0.0086$ & -5000 $-$ -1820 & $93 - 362$ & -244 $-$ -148 & $3.61 - 10.9$   \\
\hline
\hhline{|=|=|=|=|=|=|}
 $M_1$ & $M_2$ &$M_3$ &  $A_t$ & $M_{Q_3}$  & {$M_{U_3}$ }
\\
\hline 
$178 - 290$ & $304 - 5000$ & $739-5000$ & -5000 $- 142$ & $272 - 5000$ & $570-5000$
 \\
\hline
\end{tabular}
\end{center}
\caption{Range of input parameters satisfying all constraints (dimensionful parameters in GeV). }
\label{tab:1}
\end{table}

Next we present the range of masses $M_{\chi^0_3}$ for which all constraints can be satisfied simultaneously in the form of Fig.~1 showing the signal cross section $\sigma_\text{eff}^\text{signal}$ as function of $M_{\chi^0_3}$. It allows to select allowed values of $M_{\chi^0_3}$ in case more restricted values for $\sigma_\text{eff}^\text{signal}$ are preferred\footnote{The lower branch corresponds to a somewhat larger loop induced BR$(\chi^0_3\to \chi^0_2 + \gamma)$, which reduces the BR$(\chi^0_3\to \chi^0_2 + Z^*)$ required for the signal.}.
Subsequently, Fig.~2 allows one to deduce the values for $M_{\chi^\pm_1}$ and $M_{\chi^0_1}$ from the range of $M_{\chi^0_3}$.

\begin{figure}[ht!]
\begin{center}
\includegraphics[scale=0.4]{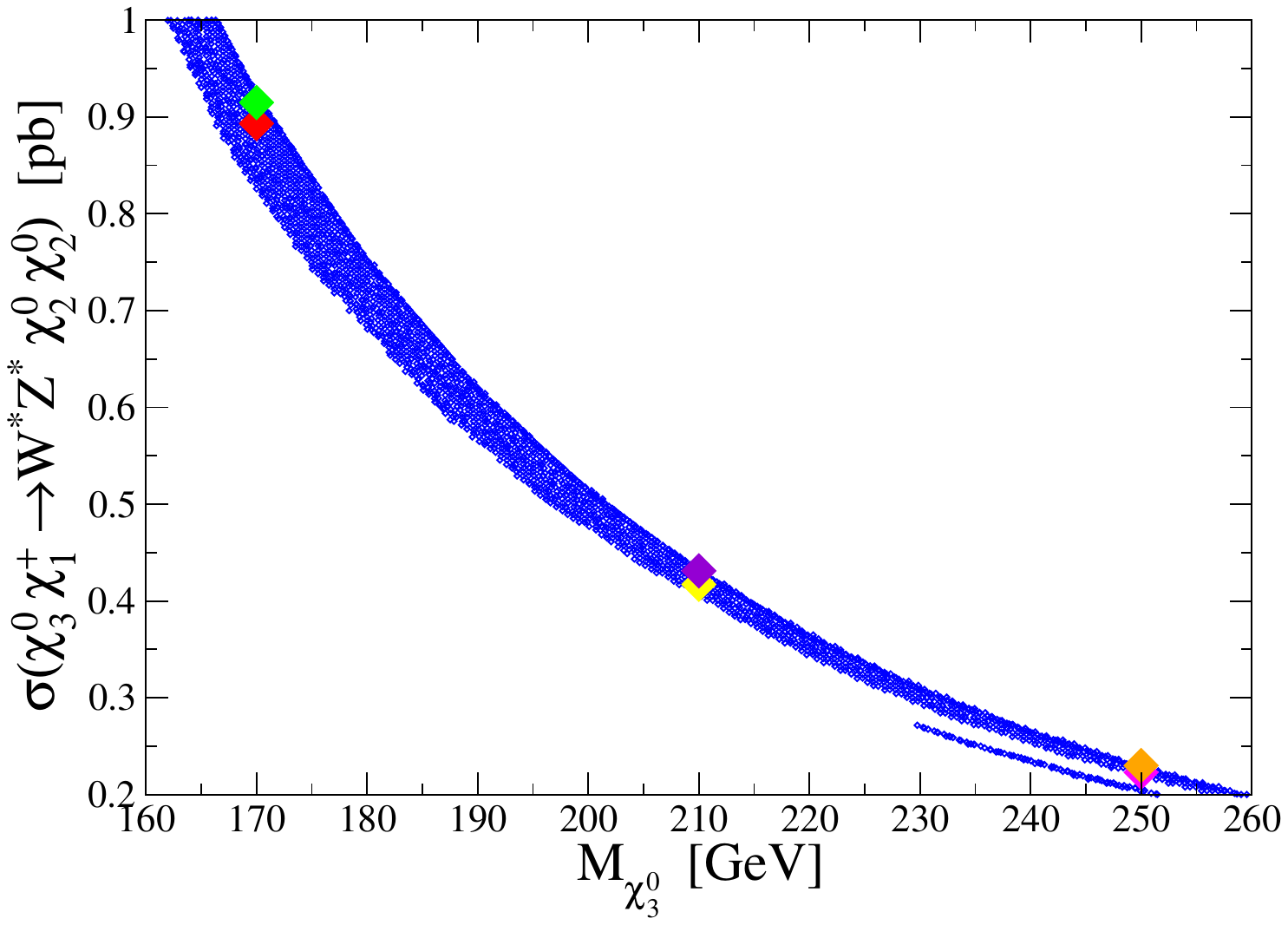}
\end{center}
\caption{The values of $\sigma_\text{eff}^\text{signal}$ for points in the NMSSM satisfying all constraints  as function of $M_{\chi^0_3}$. 
The coloured dots denote benchmark points whose details are given in the Tables~2-4 below.}
\label{fig:1}
\end{figure}

Fig.~3 shows  the points satisfy the constraint \eqref{delta_mneu} on the masses of the higgsinos as function of $\Delta m = M_{\chi_3^0}- M_{\chi_2^0}$ defined in \eqref{delta_m}. We see that even a value close to 0 would not be difficult to achieve in case this gives the best fit to the data.
Again, as function of $\Delta m$, we show in Fig.~3 also the mass difference between the higgsino NLSP and  singlino LSP, $M_{\chi^0_2} - M_{\chi^0_1}$. This mass difference plays the role of $\varepsilon$ in eq.~\eqref{dm2m1}.

Fig.~4 show the signal rates $\mu^{\rm LEP}_{bb}$ (\ref{mulep}) and $\mu_{\gamma\gamma}^{\rm LHC}$ (\ref{mugamgam}) as function of $M_{\chi^0_3}$. Whereas $\mu^{\rm LEP}_{bb}$ is obtained within the $1\,\sigma$ level $\mu_{\gamma\gamma}^{\rm LHC}$ is obtained only within the $2\,\sigma$ level.

\begin{figure}[t!]
\vspace*{-2.0cm}
\begin{center}
\hspace*{-5mm}
\begin{tabular}{cc}
\includegraphics[scale=0.3]{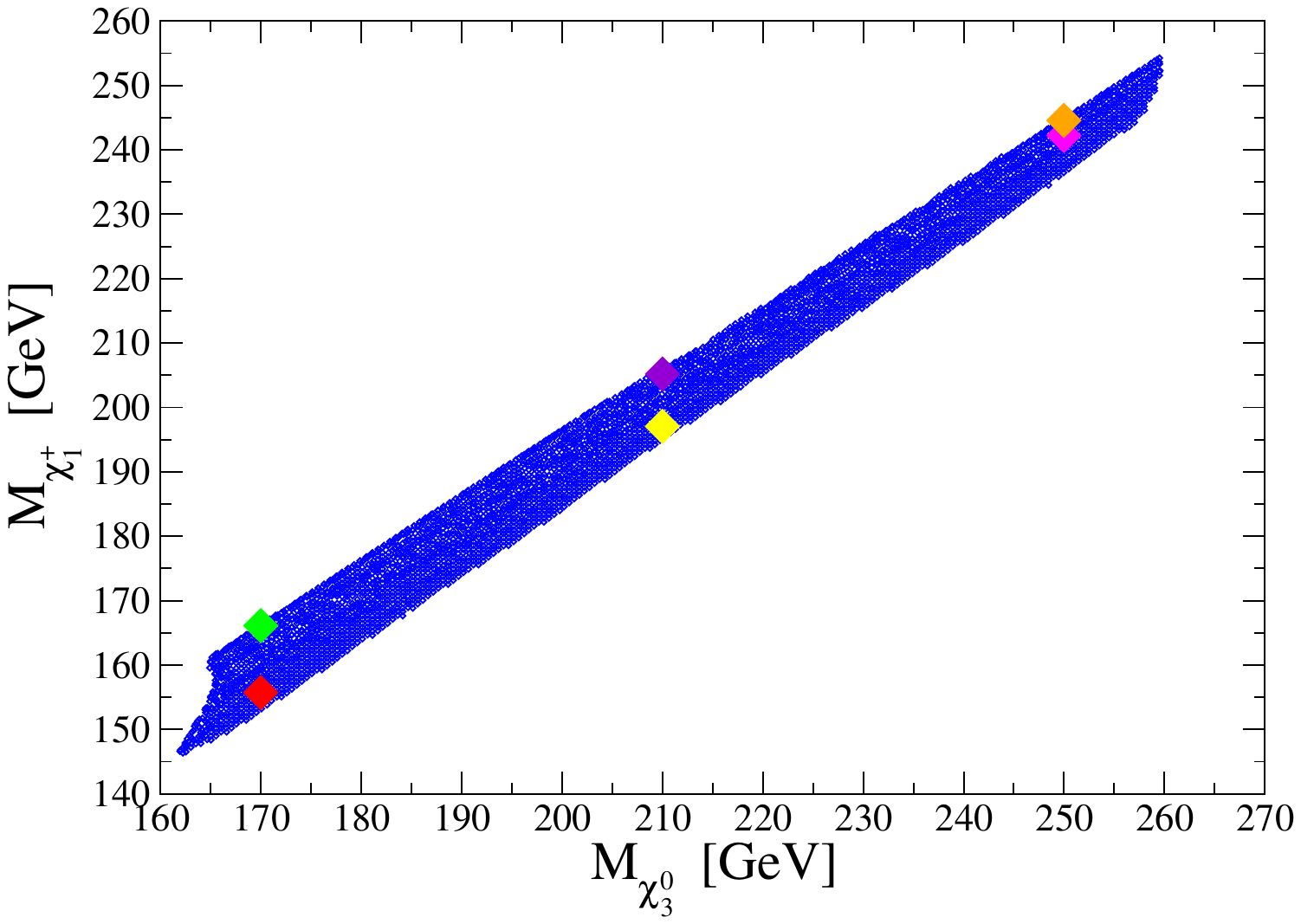}
   & 
\includegraphics[scale=0.3]{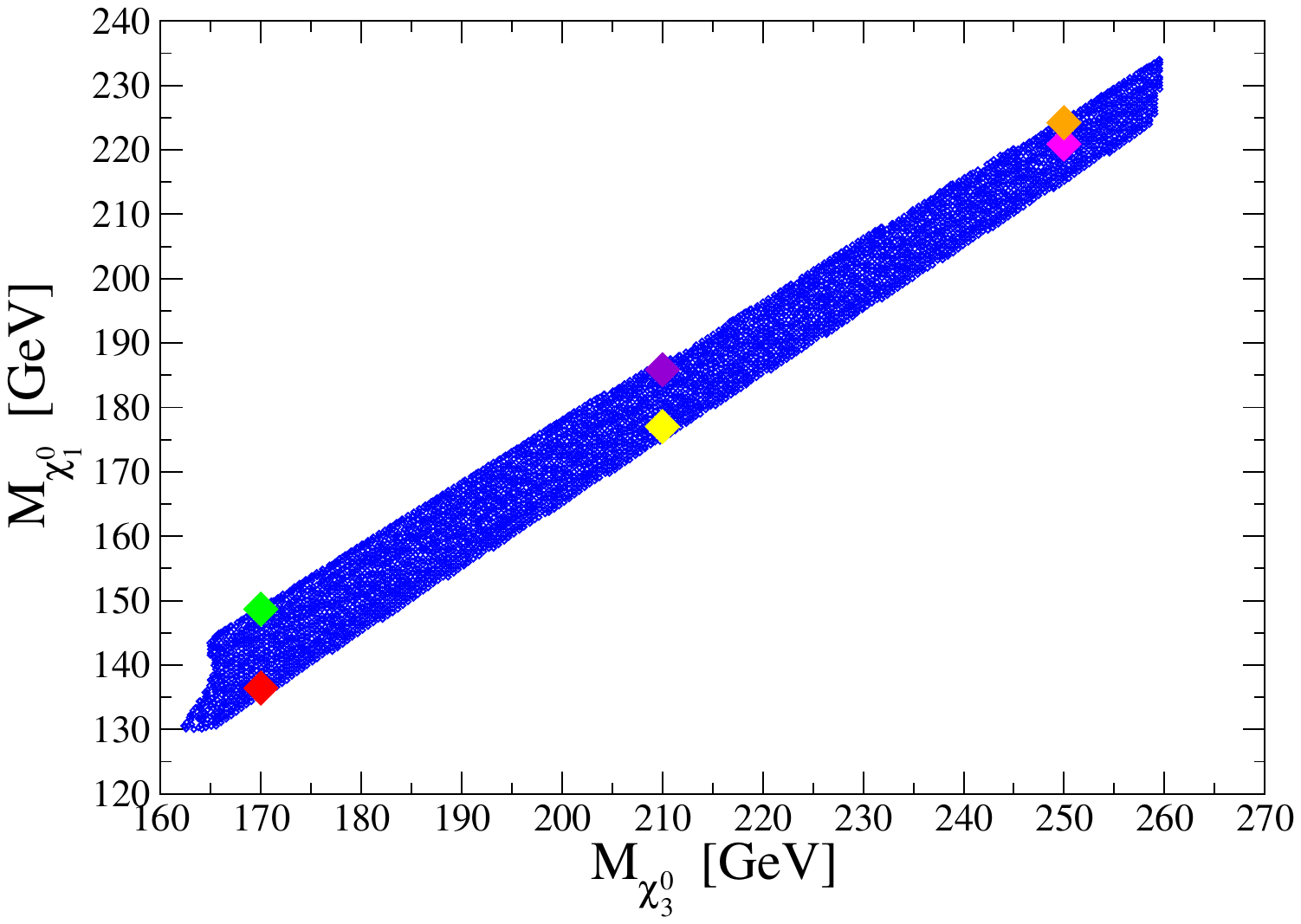}
\end{tabular}
\end{center}
\caption{$M_{\chi^\pm_1}$ (left) and $M_{\chi^0_1}$ (right) as function of of $M_{\chi^0_3}$. }
\label{fig:2}
\end{figure}

\begin{figure}[h!]
\vspace*{-0.5cm}
\begin{center}
\hspace*{-5mm}
\begin{tabular}{cc}
\includegraphics[scale=0.3]{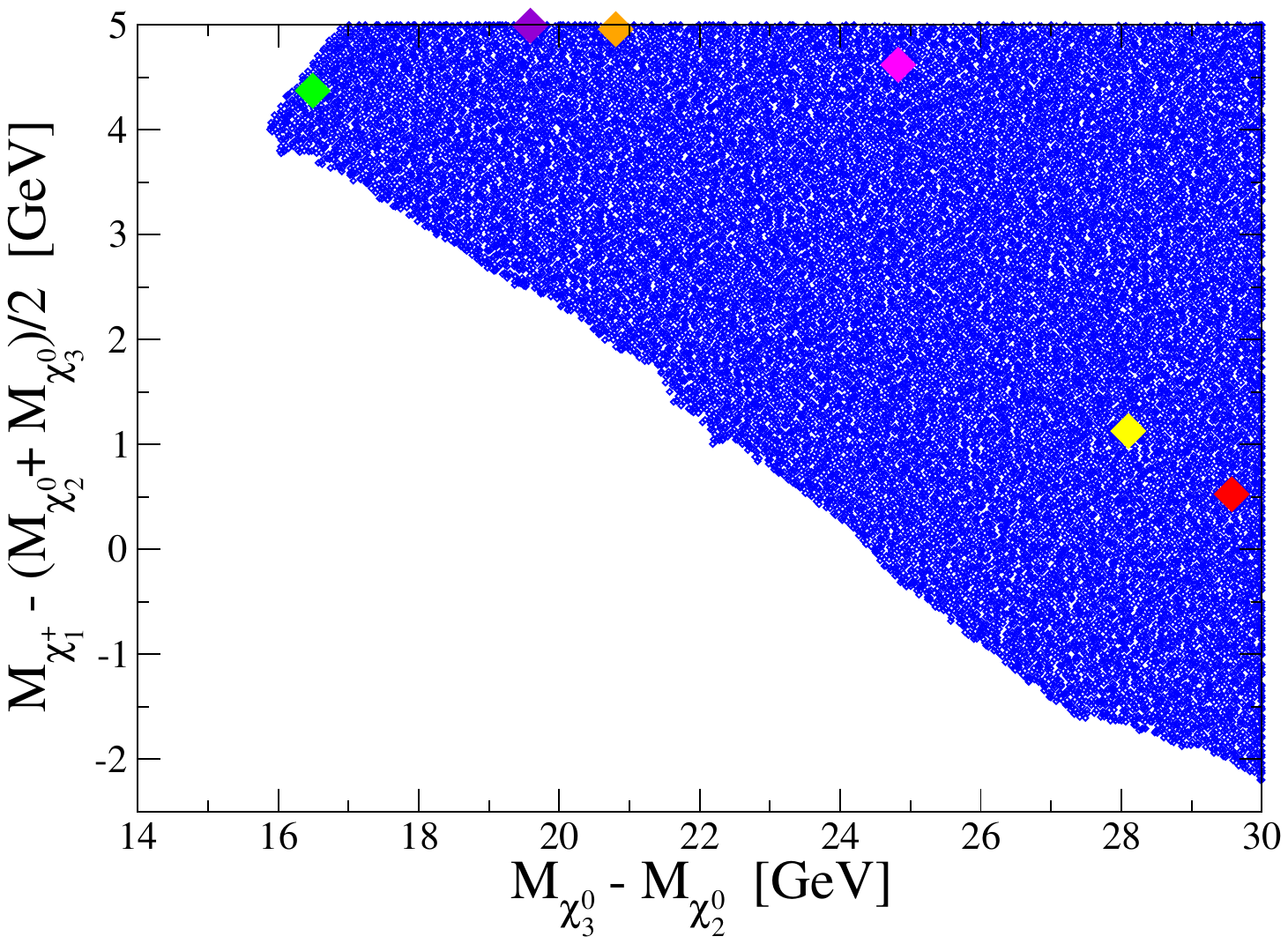}
   & 
\includegraphics[scale=0.3]{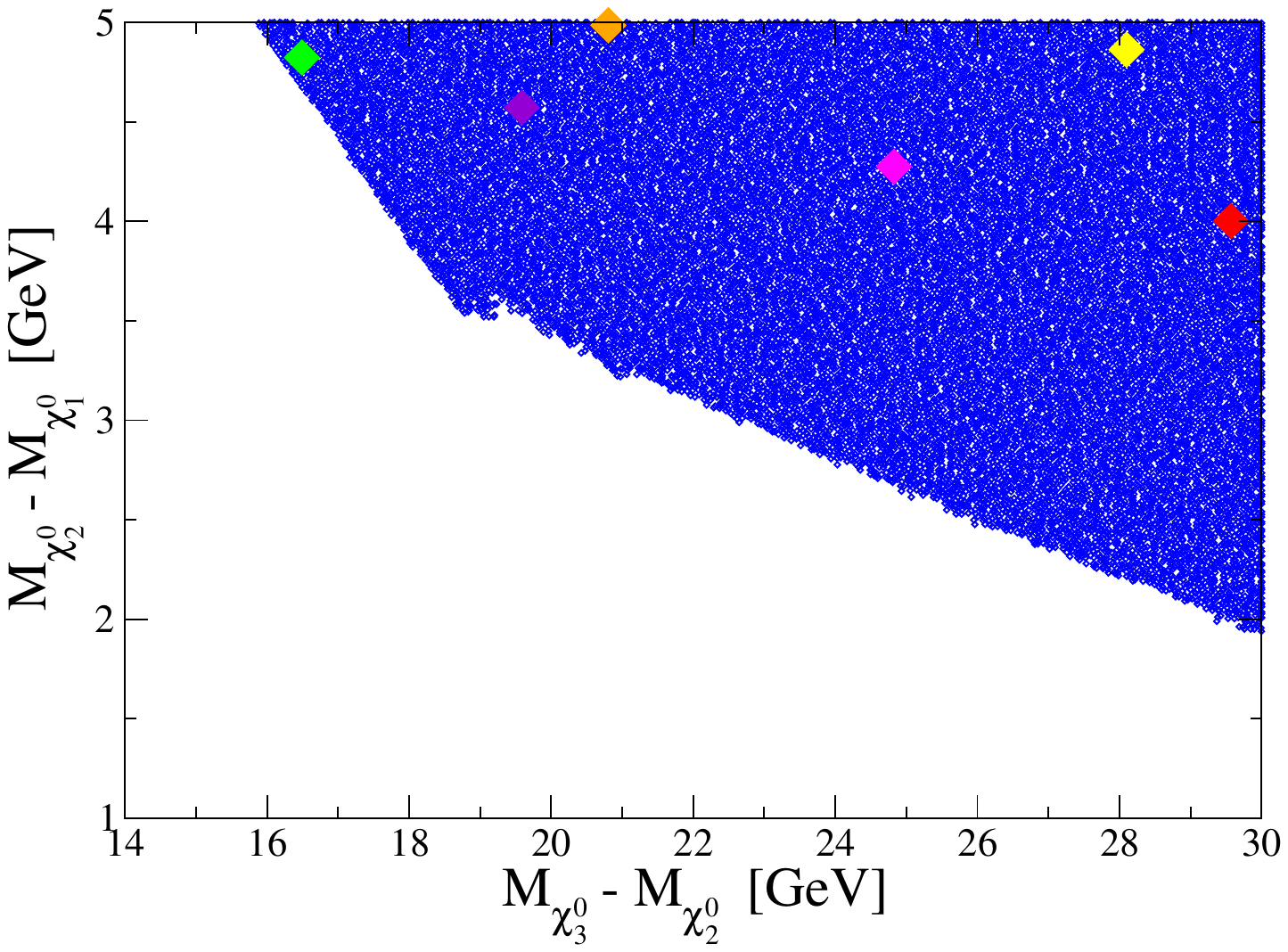}
\end{tabular}
\end{center}
\caption{The constraint \eqref{delta_mneu} on the mass of the lighter chargino and the
mass difference between higgsino NLSP and the singlino LSP as function of $\Delta m = M_{\chi_3^0}- M_{\chi_2^0}$.}
\label{fig:3}
\end{figure}

\begin{figure}[h!]
\vspace*{-0.5cm}
\begin{center}
\hspace*{-5mm}
\begin{tabular}{cc}
\includegraphics[scale=0.3]{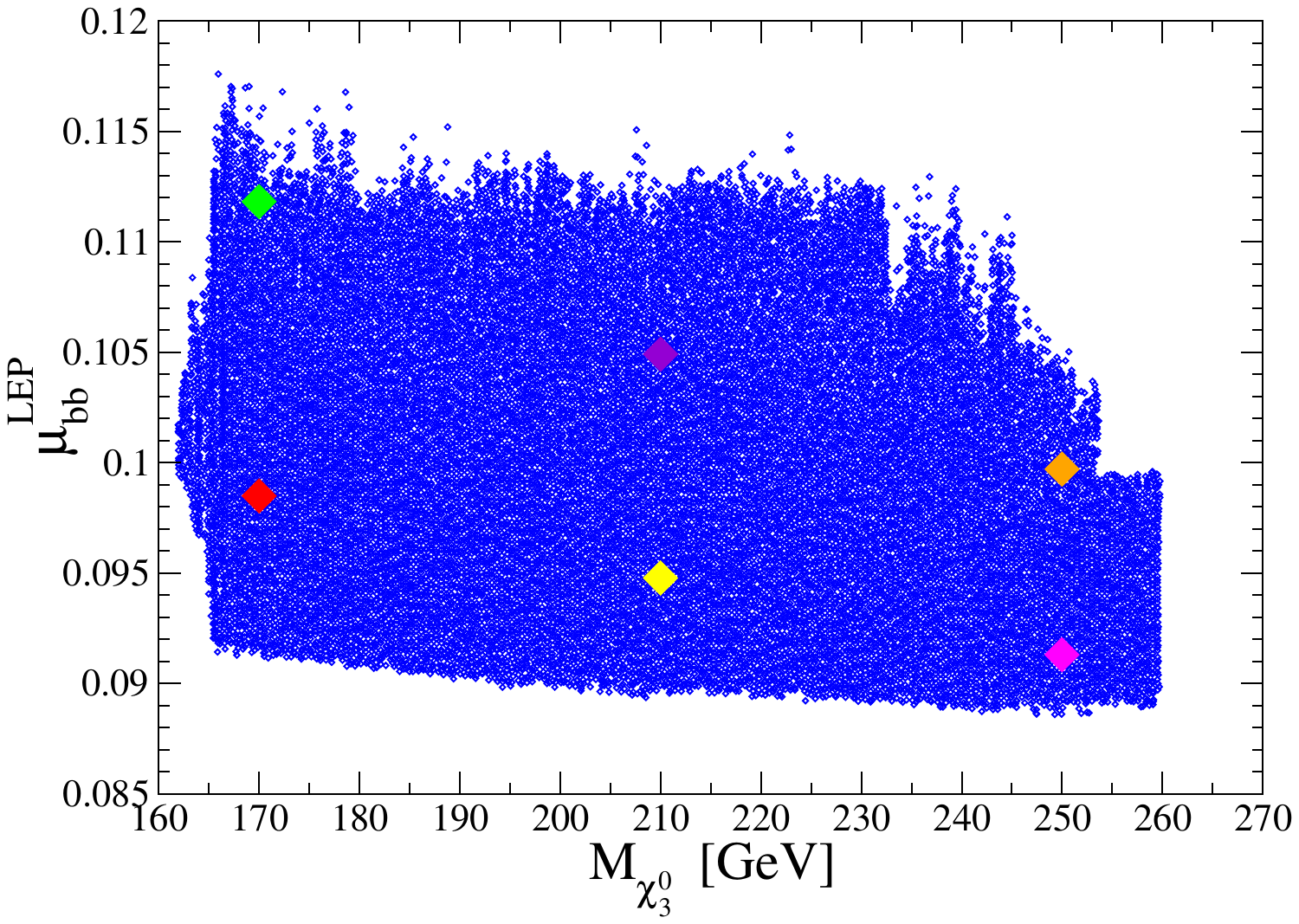}
   & 
\includegraphics[scale=0.3]{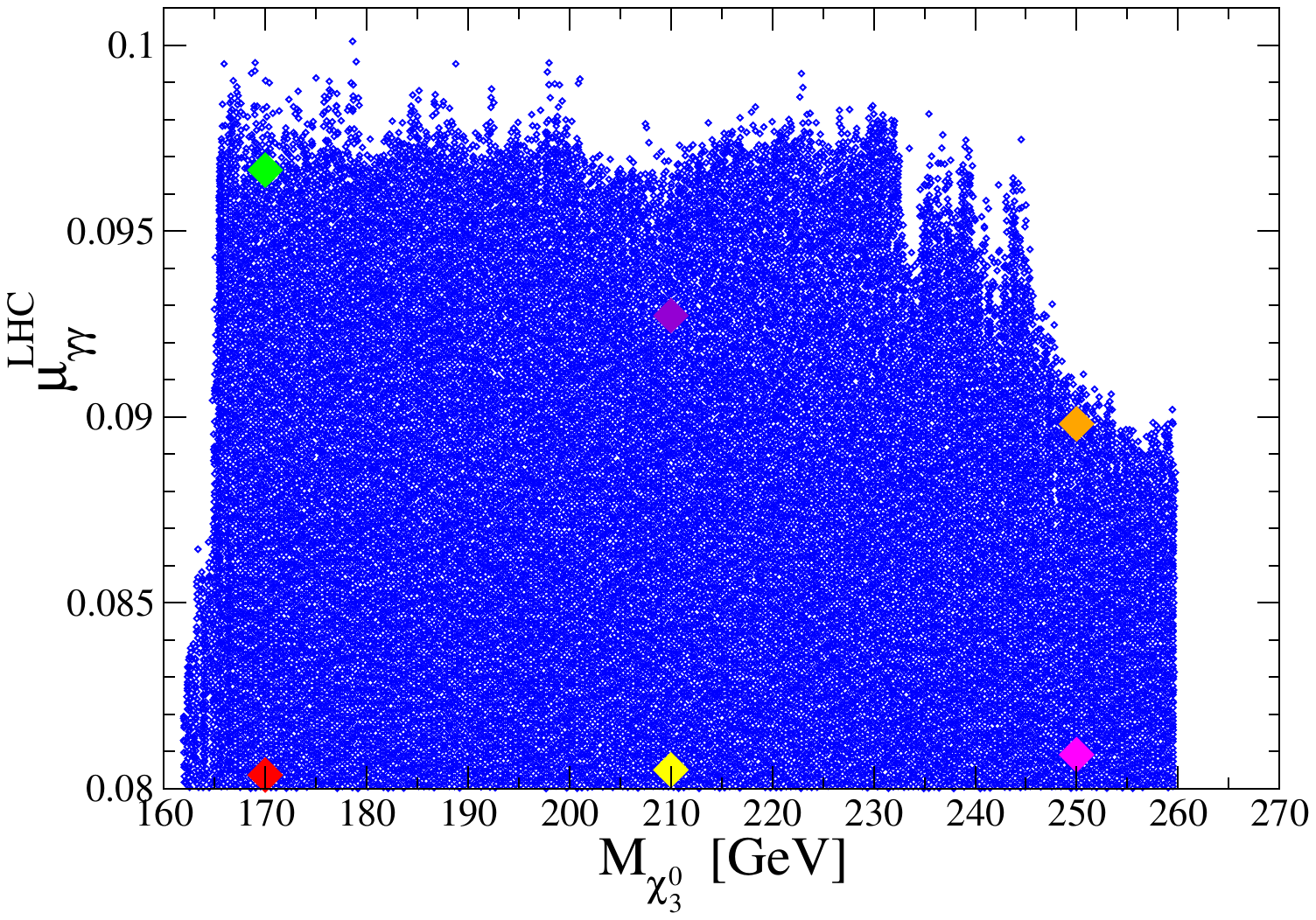}
\end{tabular}
\end{center}
\caption{$\mu^{\rm LEP}_{bb}$ from (\ref{mulep}) (left) and $\mu_{\gamma\gamma}^{\rm LHC}$ from (\ref{mugamgam}) (right) as function of $M_{\chi^0_3}$.}
\label{fig:4}
\end{figure}

\clearpage

Fig.~5 show the DM spin-idependent (SI) and spin-dependent (SD) direct 
detection rates $\sigma_{p}^{\rm SI}$ and $\sigma_{n}^{\rm SD}$ as function of $M_{\chi^0_3}$ together with the projected sensitivity of the LZ experiment from \cite{LZ:2018qzl} in red and the neutrino floor from \cite{Billard:2013qya} in green. Note that present constraints are satisfied automatically due to the singlet-nature of the LSP: we find a few points within the projected sensitivity of the LZ experiment but all points are above the neutrino floor.

\begin{figure}[ht!]
\begin{center}
\hspace*{-5mm}
\begin{tabular}{cc}
\includegraphics[scale=0.3]{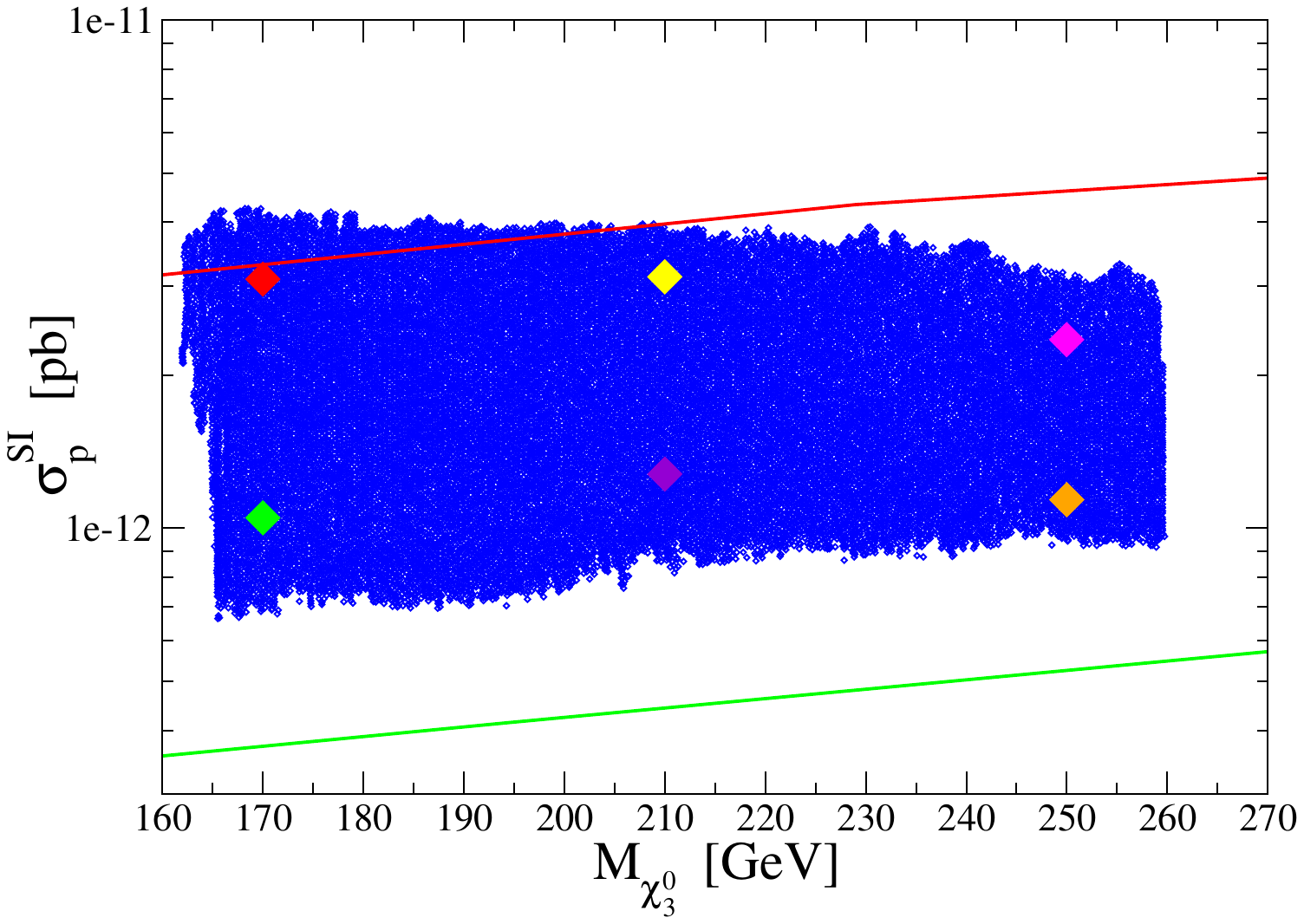}
   & 
\includegraphics[scale=0.3]{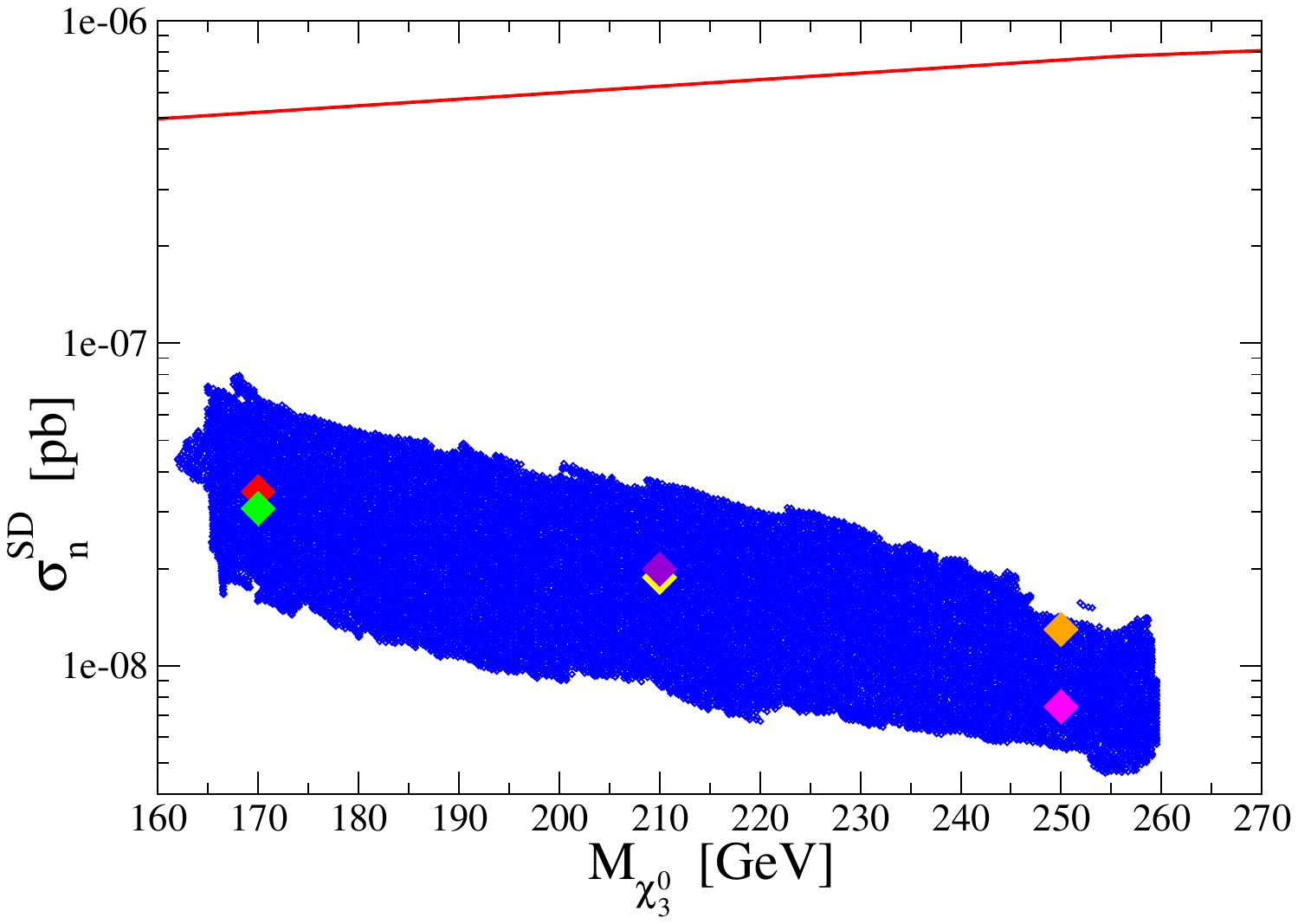}
\end{tabular}
\end{center}
\caption{$\sigma_{p}^{\rm SI}$ (left) and $\sigma_{n}^{\rm SD}$ (right) as function of $M_{\chi^0_3}$, together with the projected sensitivity of the LZ experiment from \cite{LZ:2018qzl} in red and the neutrino floor from \cite{Billard:2013qya} in green.}
\label{fig:5}
\end{figure}

The coloured dots in all figures indicate six benchmark points: BP1 (red), BP2 (green), BP3 (yellow), BP4 (violet), BP5 (pink) and BP6 (orange).
In Tables 2-4 we show their properties: input parameters in Table~2 and lightest neutralino and chargino masses  in Table~3 while 
signal rates  of the benchmark points shown in Figs.~4-5 are given in Table~4.

\begin{table} [ht!]
\begin{center}
\begin{tabular}{| c | c | c | c | c | c | c |}
\hline
  & BP1  & BP2 & BP3 & BP4  & BP5 & BP6 \\
\hline
$\lambda$        & $1.74 \!\times\! 10^{-2}$ & $1.44 \!\times\! 10^{-2}$ & $1.71 \!\times\! 10^{-2}$ & $1.52 \!\times\! 10^{-2}$ & $1.44 \!\times\! 10^{-2}$ & $1.53 \!\times\! 10^{-2}$ \\
\hline
$\kappa$          & $7.48 \!\times\! 10^{-3}$ & $6.48 \!\times\! 10^{-3}$ & $7.57 \!\times\! 10^{-3}$ & $6.93 \!\times\! 10^{-3}$ & $6.58 \!\times\! 10^{-3}$ & $7.04 \!\times\! 10^{-3}$ \\
\hline
$A_\lambda$    & -4287 & -3314 & -3624 & -3551 & -3416 & -4128 \\
\hline
$A_\kappa$      & 124 & 149 & 240 & 252 & 339 & 336 \\
\hline
$\mu_{\rm eff}$ & -155 & -161 & -197 & -200 & -234 & -235 \\
\hline
$\tan\beta$       & 8.35 & 8.81 & 6.33 & 6.30 & 5.28 & 7.19 \\
\hline
$M_1$              & 210 & 207 & 246 & 242 & 271 & 282 \\
\hline
$M_2$              & 353 & 4853 & 391 & 1296 & 597 & 964 \\
\hline
$M_3$              & 3822 & 3616 & 2264 & 2444 & 832 & 2259 \\
\hline
\end{tabular}
\caption{NMSSM specific input parameters and bino ($M_1$), wino ($M_2$) and gluino ($M_3$) masses for the six benchmark points. All dimensionful parameters are given in GeV.}
\end{center}
\label{tab:2}
\end{table}

\begin{table}[ht!]
\begin{center}
\begin{tabular}{| c | c | c | c | c | c | c |}
\hline
          & BP1  & BP2 & BP3 & BP4  & BP5 & BP6 \\
\hline
$M_{\chi^0_1}$    &136 &149 & 177 & 186 & 221 & 224 \\
\hline
$M_{\chi^0_2}$    & 140 & 154 & 182 & 190 & 225 & 229 \\
\hline
$M_{\chi^0_3}$    & 170 & 170 & 210 & 210 & 250 & 250 \\
\hline
$M_{\chi^\pm_1}$ & 156 & 166 & 187 & 205 & 242 & 245 \\
\hline
\end{tabular}
\caption{Lightest neutralino and chargino masses (in GeV)  for the six benchmark points.}
\end{center}
\label{tab:3}
\end{table}

\begin{table}[ht!]
\begin{center}
\begin{tabular}{| c | c | c | c | c | c | c |}
\hline
  & BP1  & BP2 & BP3   & BP4  & BP5 & BP6 \\
\hline
$\mu^{\rm LEP}_{bb}$  & $9.85 \!\times\! 10^{-2}$ & 0.112 & $9.48 \!\times\! 10^{-2}$ & 0.105 & $9.13 \!\times\! 10^{-2}$ & $9.97 \!\times\! 10^{-2}$  \\
\hline
$\mu^{\rm LHC}_{\gamma\gamma}$ & $8.04 \!\times\! 10^{-2}$ & $9.66 \!\times\! 10^{-2}$ & $8.05 \!\times\! 10^{-2}$ & $9.27 \!\times\! 10^{-2}$ & $8.09 \!\times\! 10^{-2}$ & $8.98 \!\times\! 10^{-2}$ \\
\hline
$\sigma_{p}^{\rm SI}$ & $3.09 \!\times\! 10^{-12}$ & $1.05 \!\times\! 10^{-12}$ & $3.13 \!\times\! 10^{-12}$ & $1.28 \!\times\! 10^{-12}$ & $2.35 \!\times\! 10^{-12}$ & $1.14 \!\times\! 10^{-12}$\\
\hline
$\sigma_{n}^{\rm SD}$ & $3.47 \!\times\! 10^{-8}$ & $3.07 \!\times\! 10^{-8}$ & $1.88 \!\times\! 10^{-8}$ & $1.99 \!\times\! 10^{-8}$ & $7.46 \!\times\! 10^{-9}$ & $1.29 \!\times\! 10^{-8}$\\
\hline
\end{tabular}
\caption{The signal rates $\mu^{\rm LEP}_{bb}$, $\mu^{\rm LHC}_{\gamma\gamma}$ and DM direct detection cross sections $\sigma_{p}^{\rm SI}$ and $\sigma_{n}^{\rm SD}$ (the latter in pb) for the six benchmark points.}
\end{center}
\label{tab:4}
\end{table}


\section{Conclusions}

Although the observed excesses in the search for neutralinos and charginos by ATLAS and CMS can in principle be interpreted in the MSSM, assuming a light higgsino mass and a {\em compressed} higgsino dominated neutralino and chargino spectrum, such light higgsinos as DM would have far too large direct detection cross sections. 
In this paper we have instead considered the NMSSM with an additional singlino LSP a few GeV below the NLSP. Herein, sparticles predominantly decay first into the NLSP and remnants from the final decay into the LSP are too soft to contribute to the observed signals. 
In contrast to the MSSM, a singlino LSP in the NMSSM allows to describe easily a DM relic density in the WMAP/Planck window and is consistent with present limits on the DM direct detection cross sections. 

The considered NMSSM scenario consists of the ``compressed'' light higgsino-like triplet made up by   ($\chi_3^0$, $\chi_1^{\pm}$, $\chi_2^0$)  plus the nearby singlino LSP $\chi_1^0$, with all masses less than about 250 GeV with of order $5-10\%$ mass splittings. 
In this scenario, if the LSP is very close in mass to the NLSP, co-annihilation in the higgsino sector can generate a relic density in the WMAP/Planck window. This allows us to describe the excesses observed by the above ATLAS and CMS searches together with DM which automatically satisfies present (and even future) limits on the DM direct detection cross sections. Indeed, our scenario can be falsified by the observation of direct detection cross sections. 

Simultaneously, the additional singlet-like Higgs scalar can describe excesses in the $b\bar{b}$ channel at LEP and in the $\gamma\gamma$ (but not $\tau\tau$) channel at the LHC at around 95~GeV, with signal cross sections compatible with the respective observations. Clearly, these and the excesses in the search for neutralinos and charginos remain to be confirmed. If this is the case, the NMSSM would provide a very promising framework for their common explanation.

Finally, to invite exploration of possible solutions provided by this SUSY framework, we have defined six benchmark points amenable to further phenomenological investigation.  

\section*{Acknowledgements}
SFK would like to thank CERN for hospitality and acknowledges the STFC Consolidated Grant ST/L000296/1 and the European Union's Horizon 2020 Research and Innovation programme under Marie Sklodowska-Curie grant agreement HIDDeN European ITN project (H2020-MSCA-ITN-2019//860881-HIDDeN). SM is supported in part through the NExT Institute and the  STFC Consolidated Grant ST/L000296/1.

\clearpage

\end{document}